\documentclass[
aps,
prx,
reprint,
superscriptaddress,
longbibliography
]{revtex4-2}

\usepackage{graphicx,physics}% Include figure files
\usepackage{color}
\usepackage{dcolumn,comment}% Align table columns on decimal point
\usepackage{latexsym}

\usepackage[normalem]{ulem}
\usepackage{hyperref,amssymb}
\usepackage{url}
\usepackage{color,soul,xcolor}
\usepackage{graphicx}
\usepackage{verbatim}
\usepackage{multirow}
\usepackage{amsmath}
\newcommand{\beq}{\begin{eqnarray}}
\newcommand{\eeq}{\end{eqnarray}}
\usepackage{mathrsfs}
\usepackage{float,soul}
\usepackage[dvipsnames]{xcolor}
\usepackage{mathtools}
\usepackage{slashed}
\usepackage{physics}	% installed packages for typing maths and physics
\usepackage{graphicx} % need for Fig.ures
\usepackage{epstopdf}
\usepackage{subfigure} % use for side-by-side Fig.ures
\usepackage{bbold}
\usepackage{wasysym}
\usepackage{feynmp}
\usepackage{hyperref}
\usepackage{tabularx}
\usepackage{placeins}
\usepackage{booktabs}
\usepackage[table]{xcolor}

\definecolor{headergray}{RGB}{236,239,243}
\definecolor{rowgray}{RGB}{248,249,251}
\hypersetup{colorlinks,
}
\usepackage{bm}%varepsilon

\usepackage{cancel}

\begin{document}

\title{The End of Chaos at the Ideal Glass Transition}

\author{Zi-Long Wang}
\affiliation{State Key Laboratory of Polymer Science and Technology, Changchun Institute of Applied Chemistry, Chinese Academy of Sciences, Changchun 130022, P. R. China}
\affiliation{School of Applied Chemistry and Engineering, University of Science and Technology of China, Hefei 230026, P. R. China}

\author{Jack F. Douglas}
\email{jack.douglas@nist.gov}
\affiliation{Materials Science and Engineering Division, National Institute of Standards and Technology, Gaithersburg, Maryland 20899, United States}

\author{Matteo Baggioli}
\email{b.matteo@sjtu.edu.cn}
\affiliation{Wilczek Quantum Center, School of Physics and Astronomy, Shanghai Jiao Tong University, Shanghai 200240, P. R. China}
\affiliation{Shanghai Research Center for Quantum Sciences, Shanghai 201315, P. R. China}

\author{Zhao-Yan Sun}
\email{zysun@ciac.ac.cn}
\affiliation{State Key Laboratory of Polymer Science and Technology, Changchun Institute of Applied Chemistry, Chinese Academy of Sciences, Changchun 130022, P. R. China}
\affiliation{School of Applied Chemistry and Engineering, University of Science and Technology of China, Hefei 230026, P. R. China}
\affiliation{Jilin Provincial International Cooperation Key Laboratory for Polymer Processing Physics, Changchun 130022, P. R. China}

\author{Wen-Sheng Xu}
\email{wsxu@nju.edu.cn}
\affiliation{MOE Key Laboratory of High Performance Polymer Materials and Technology, State Key Laboratory of Coordination Chemistry, School of Chemistry, Nanjing University, Nanjing 210023, P. R. China}

\begin{abstract}
The ideal glass transition has often been associated with the extrapolated vanishing of configurational entropy $S_c$, yet how this thermodynamic criterion is reflected in microscopic dynamics remains unclear. Here, we show that the vanishing of deterministic chaos quantified by the maximum Lyapunov exponent $\lambda_{\mathrm{max}}$ provides such a dynamical signature. Using molecular dynamics simulations across diverse model glass-forming liquids in two and three dimensions, we show that $\lambda_{\mathrm{max}}$ follows an approximate scaling $\lambda_{\mathrm{max}} \propto (T-T_{\lambda})^{1/2}$ and extrapolates to zero at a temperature $T_{\lambda}$ consistent with the Kauzmann temperature $T_{\mathrm{K}}$. Furthermore, we establish the quantitative relation $\lambda_{\mathrm{max}}^2 \propto T S_c$, directly linking chaotic dynamics to configurational entropy. By demonstrating that $\lambda_{\mathrm{max}}^2 \propto \sigma_K^2$, where $\sigma_K^2$ is the variance of local potential energy landscape curvature fluctuations, we connect the suppression of chaos directly to landscape geometry. These results provide a unified thermodynamic, dynamical, and geometric framework for glass formation, identifying the ideal glass transition as the endpoint of chaotic exploration of the energy landscape.
\end{abstract}

\maketitle

The dramatic slowdown of dynamics upon approaching the glass transition remains one of the central unsolved problems in condensed-matter physics~\cite{Supercooled_2001_410_259a,Facets_2016_69_40}. Upon cooling, the structural relaxation time of a glass-forming liquid increases by many orders of magnitude, often following a super-Arrhenius temperature ($T$) dependence, despite little or no detectable change in its static structure. Understanding the microscopic origin of this extraordinary dynamical slowdown lies at the heart of glass science~\cite{Theoretical_2011_83_587a}.

One influential view is that this slowdown reflects an underlying thermodynamic transition to an ideal glass state. This scenario traces back to Kauzmann's observation~\cite{nature_1948_43_219b} that the entropy of a supercooled liquid decreases so rapidly upon cooling that its extrapolated value would eventually fall below that of the crystal, leading to an apparent entropy crisis. The proposed resolution is that the configurational entropy $S_c$, defined as the total entropy minus the vibrational contribution~\cite{Configurational_2019_150_160902}, extrapolates to zero at a finite Kauzmann temperature $T_{\mathrm{K}}$, which within this theoretical framework defines a putative ideal glass transition~\cite{kauzmann_2023__203}. The idea of an ideal thermodynamic glass transition was developed theoretically by Gibbs and DiMarzio \cite{Nature_1958_28_373} based on a lattice model of polymers, while Adam and Gibbs \cite{temperature_1965_43_139a} later connected the depletion of configurational entropy to the dramatic slowing down of structural relaxation. However, the microscopic dynamical mechanism underlying this connection remains unclear: the configurational entropy quantifies how many amorphous states are available, but does not specify how the ability of the system to dynamically explore them is progressively lost.

A compelling theoretical realization of this scenario emerges in the limit of infinite dimensions, where mean-field theory becomes exact~\cite{Meanfield_2010_82_789,Glass_2017_8_265}. Within this approximation, random first-order transition (RFOT) theory predicts a genuine thermodynamic transition at a finite temperature $T_{\mathrm{K}}$, where the configurational entropy vanishes and an overlap order parameter distinguishing amorphous states becomes non-zero~\cite{Theory_2007_58_235,Colloquium_2015_87_183}. Whether this elegant mean-field picture survives in finite dimensions remains one of the most intensely debated questions in glass physics.

\begin{figure*}[ht]
	\centering
	\includegraphics[width=0.93\linewidth]{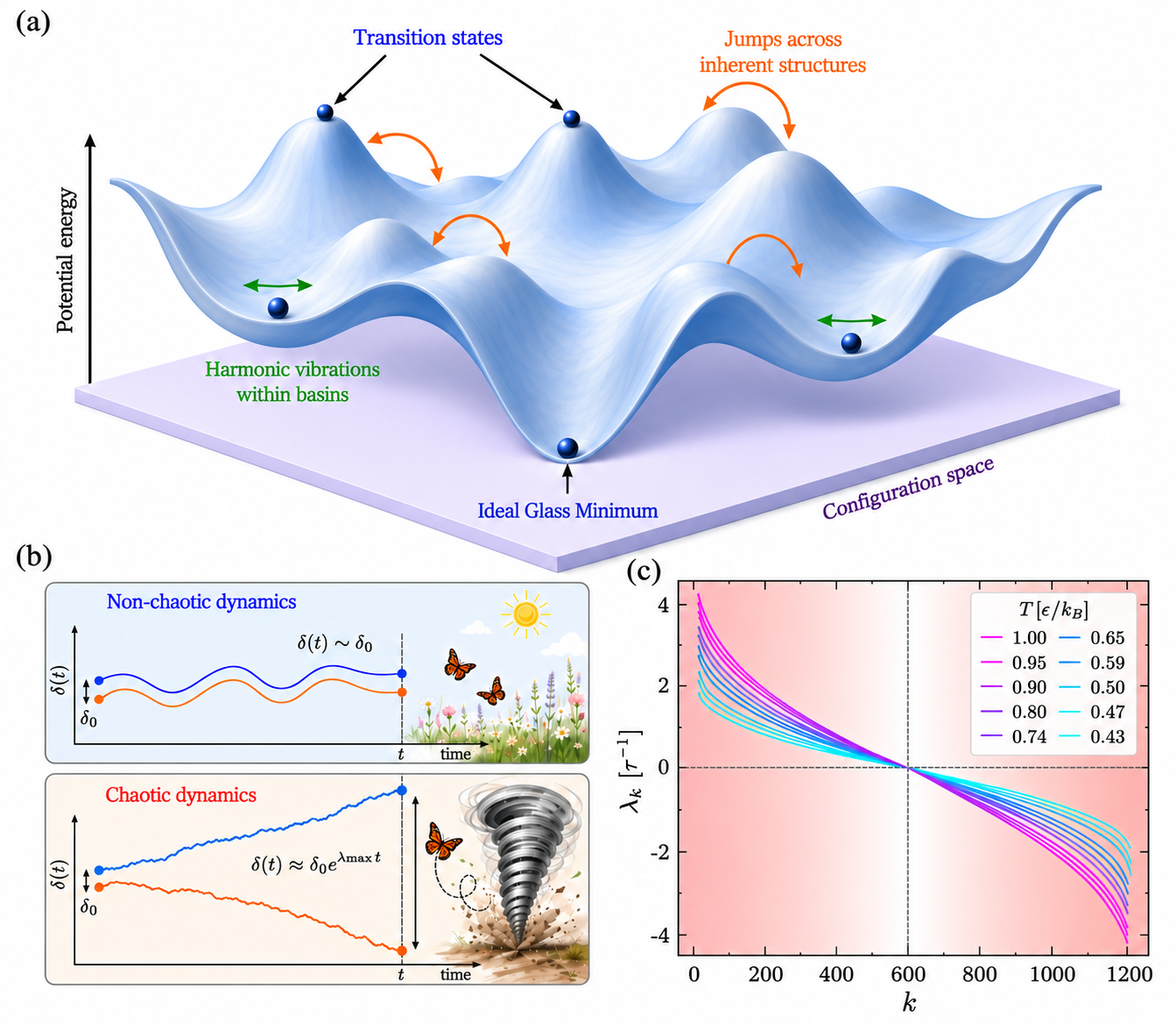}
	\caption{\textbf{Chaotic dynamics across the glass potential energy landscape (PEL).} \textbf{(a)} Schematic illustration of the PEL. The dynamics consist of a combination of quasi-harmonic vibrations within individual basins and activated transitions between distinct inherent structures. \textbf{(b)} Illustration highlighting the difference between non-chaotic and chaotic dynamics. In the latter, an infinitesimal initial separation $\delta_0$ between two trajectories grows exponentially with time $t$, $\delta(t) \sim \delta_0 e^{\lambda_{\mathrm{max}} t}$, illustrating the butterfly effect, where $\lambda_{\mathrm{max}}$ is the maximum Lyapunov exponent. \textbf{(c)} Lyapunov spectrum of the 3DKA12 model with a particle number of $N=200$ over a broad range of temperatures $T$, where $\lambda_k$ denotes the $k$th Lyapunov exponent. The smaller system is simulated only for the illustration of the Lyapunov spectrum. Panels (a) and (b) were generated with the assistance of AI-based tools.}
	\label{fig1}
\end{figure*}

Yet, even if an ideal glass transition exists, its physical nature remains elusive. Thermodynamic theories characterize the transition through the vanishing of the configurational entropy, whereas dynamical approaches emphasize the dramatic growth of relaxation times, dynamical heterogeneity, and cooperative motion~\cite{Complex_2008_143_a,Dynamics_2010_61_191a}. Despite their individual successes, these two perspectives remain only indirectly connected, and no widely accepted microscopic dynamical quantity is known to become critical at the Kauzmann temperature. Identifying such a quantity would provide a dynamical counterpart to the entropy crisis and a microscopic basis for the connection between configurational entropy and structural relaxation.

From the perspective of the potential energy landscape (PEL), chaos provides a natural candidate to fill this gap~\cite{Chaos_1998__,Chaos_2002__}. While the configurational entropy quantifies the multiplicity of amorphous states, chaotic dynamics characterize the sensitivity and instability of trajectories exploring the landscape. This behaviour is encoded in the Lyapunov spectrum~\cite{Probleme_1907_2eserie9_203}, which measures the exponential divergence of nearby trajectories and is intimately connected to the local geometry of the underlying energy landscape.

Here, we show that the loss of chaos provides the missing dynamical signature of the ideal glass transition. Across several glass-forming liquids in two and three dimensions, the maximum Lyapunov exponent $\lambda_{\mathrm{max}}$ characterizing chaotic dynamics extrapolates continuously to zero at a characteristic temperature $T_\lambda$ that closely tracks the independently determined Kauzmann temperature $T_{\mathrm{K}}$, following
$\lambda_{\mathrm{max}}\propto(T-T_\lambda)^{1/2}$ to a good approximation.
We further establish the relations
$\lambda_{\mathrm{max}}^2\propto T S_c$ and
$\lambda_{\mathrm{max}}^2\propto\sigma_K^2$,
where $\sigma_K^2$ denotes the variance of curvature fluctuations of the PEL. These results connect configurational thermodynamics, chaotic dynamics, and landscape geometry, and identify the approach to the ideal glass state with a progressive loss of the dynamical instability required to explore the energy landscape.

\begin{figure*}[htp]
	\centering
	\includegraphics[width=0.98\linewidth]{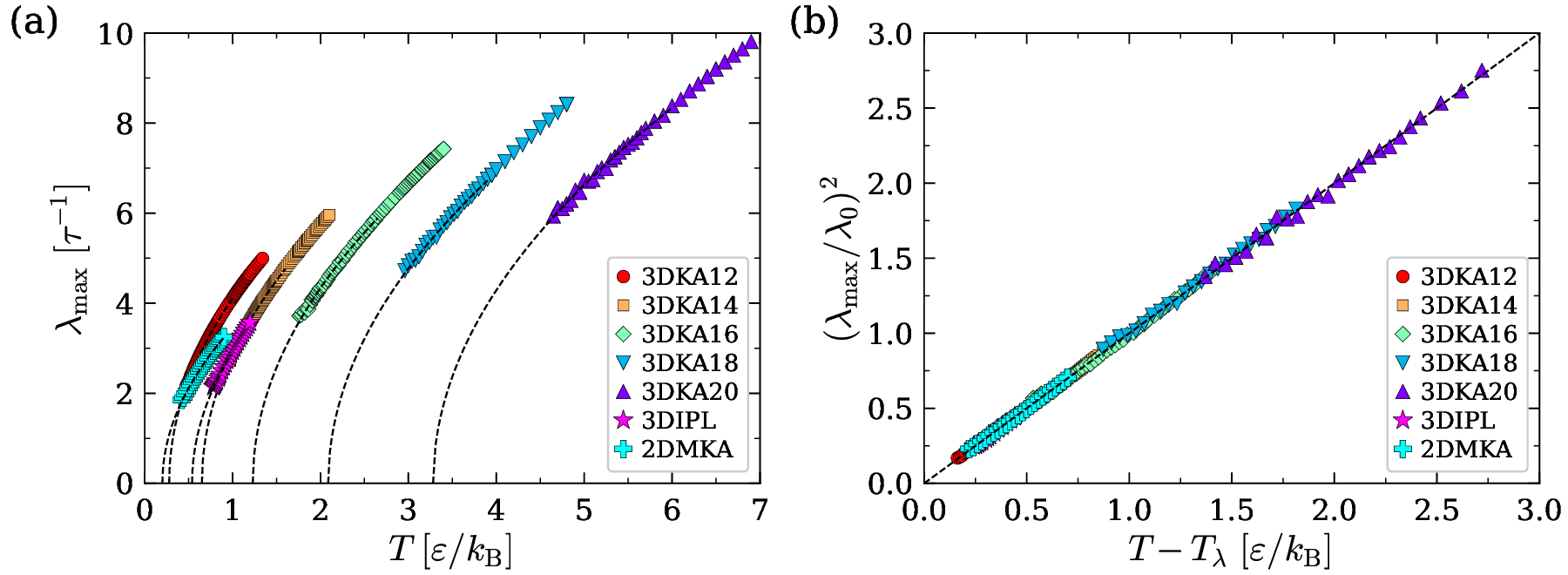}
	\caption{\textbf{Vanishing of chaotic dynamics near the ideal glass transition.} \textbf{(a)} Temperature dependence of the maximum Lyapunov exponent $\lambda_{\mathrm{max}}$ for the two- and three-dimensional glass-forming systems considered. The notation 3DKA$\#$ denotes the three-dimensional Kob--Andersen system with $\#$ indicating the number density $\rho$. Black dashed curves show extrapolations using $\lambda_{\mathrm{max}} = \lambda_0 (T-T_\lambda)^{1/2}$; unconstrained fits with a free exponent are shown in the Supplementary Information (SI). \textbf{(b)} Reduced representation of the same data, showing the common scaling $\lambda_{\mathrm{max}}^2 \propto T - T_\lambda$ after accounting for the system-dependent amplitude $\lambda_0$.}
	\label{fig2}
\end{figure*}
\section*{From Energy Landscape to Chaos}

The PEL framework provides a natural language to connect thermodynamic and dynamical aspects of glass formation~\cite{Viscous_1969_51_3728b, topographic_1995_267_1935a, Potential_2005_2005_P05015a}. Within this picture, a supercooled liquid explores a complex landscape containing an exponentially large number of metastable basins connected through unstable directions and saddle points [Fig.~\ref{fig1}(a)]. While the configurational entropy measures the number of accessible basins, the dynamics is determined by the system's ability to navigate among them. This observation naturally raises a central question: how can one quantify the exploration of the energy landscape itself?

Chaos is a well-established concept in nonlinear dynamics that plays a fundamental role in a variety of important phenomena, ranging from weather forecasting and financial markets to many-body physical systems~\cite{Chaos_1998__,Chaos_2002__}. Broadly speaking, chaotic dynamics are characterized by the exponential amplification of an infinitesimal difference in initial conditions, causing two initially neighboring trajectories to diverge exponentially in time. This sensitive dependence on initial conditions, commonly known as the butterfly effect, is illustrated schematically in Fig.~\ref{fig1}(b).

Mathematically, the dynamics of an $N$-particle Hamiltonian system are described in phase space by $\dot{\mathbf{X}} = \mathbf{F}(\mathbf{X})$, where $\mathbf{X} = (\mathbf{r},\mathbf{p})$ collects all particle positions $\mathbf{r}$ and momenta $\mathbf{p}$ and $\mathbf{F}$ is the Hamiltonian flow. The stability of a trajectory is characterized by the evolution of an infinitesimal perturbation $\bm{\delta}(t) = [\delta\mathbf{r}(t), \delta\mathbf{p}(t)]$, which obeys the linearized dynamics $\dot{\bm{\delta}}(t) = \mathbf{J}[\mathbf{X}(t)]\bm{\delta}(t)$, where $\mathbf{J} = \partial\mathbf{F}/\partial\mathbf{X}$ is the Jacobian evaluated along the reference trajectory. For the Hamiltonians considered here, $\mathbf{J}$ is determined by the Hessian of the potential energy and therefore encodes the local curvature of the energy landscape. The long-time growth rates of linearly independent perturbations define the Lyapunov spectrum $\{\lambda_k\}$ with $k$ being the index. Additionally, the sum of the positive Lyapunov exponents defines the Kolmogorov--Sinai entropy rate $h_{\mathrm{KS}}$ that quantifies the rate at which a dynamical system generates information, or equivalently loses predictability, along its trajectory.

In this work, we performed extensive molecular dynamics simulations of representative glass-forming models spanning two and three dimensions and both attractive and purely repulsive interactions: a two-dimensional modified Kob--Andersen model (2DMKA), the canonical three-dimensional Kob--Andersen mixture at several densities (3DKA$\#$, where $\#$ is associated with the specified value of the number density $\rho$), and a three-dimensional inverse-power-law (3DIPL) system. Full details of the models and simulation protocols are provided in the \textit{Methods}. The Lyapunov spectrum is computed using the Benettin algorithm~\cite{Lyapunov_1980_15_9, Lyapunov_1980_15_21}; further computational details are provided in the \textit{Methods}.

An example of the Lyapunov spectrum is shown in Fig.~\ref{fig1}(c) for the 3DKA12 system over a range of temperatures $T$. Throughout this work, we focus on the largest Lyapunov exponent,
\begin{equation}
	\lambda_{\mathrm{max}} = \lim_{t\rightarrow\infty}\frac{1}{t} \ln\frac{\|\bm{\delta}(t)\|}{\|\bm{\delta}(0)\|},
	\label{Eq_Lmax}
\end{equation}
which measures the asymptotic exponential rate at which the most unstable perturbation grows. Thus, $\lambda_{\mathrm{max}}>0$ signals chaotic dynamics, whereas $\lambda_{\mathrm{max}} = 0$ marks the disappearance of exponential instability. In the SI, we show that $\lambda_{\mathrm{max}}$ is proportional to $h_{\mathrm{KS}}$ to a good approximation, so our findings based on $\lambda_{\mathrm{max}}$ apply to $h_{\mathrm{KS}}$ as well.

\section*{The End of Chaos at the Kauzmann Temperature}

The chaotic nature of atomic dynamics, and in particular the Lyapunov spectrum, has provided valuable insights into a broad range of problems in statistical physics, including simple fluids~\cite{Extensivity_2017_95_22102}, the jamming transition in athermal granular systems~\cite{chaotic_2013_9_288a}, liquid--vapor criticality~\cite{Critical_2019_10_2155}, relaxation and its connection to instantaneous normal modes~\cite{Lyapunov_1996_76_3738}, and crystal melting~\cite{Lyapunov_1996_230_364,Curvature_1997_56_2508}. Lyapunov measures have also been discussed in connection with relaxation in glass-forming systems (see, e.g., Refs.~\cite{Lyapunov_1996_76_3738,Crossover_2024_37_95402}). However, previous work has not established whether chaotic instability develops a critical temperature tied to the thermodynamic ideal-glass scale, nor a direct quantitative relation between the Lyapunov dynamics and configurational entropy. Accessing this regime is especially challenging because equilibration of deeply supercooled liquids and the accompanying long-time tangent-space integrations are both computationally demanding.

\begin{table}[htp]
\vspace{0.4cm}
\centering
\renewcommand{\arraystretch}{1.35}
\setlength{\tabcolsep}{10pt}

\caption{\textbf{Comparison of characteristic temperatures.} The Kauzmann temperature $T_{\mathrm{K}}$ is obtained by extrapolating the configurational entropy to zero, $T_0$ denotes the Vogel--Fulcher--Tammann (VFT) temperature extracted from fits of the structural relaxation time, and $T_\lambda$ is obtained from the low-temperature scaling of the maximum Lyapunov exponent.}
\label{tab:Tk}

\begin{tabular}{lccc}
\toprule
\textbf{Model} & \textbf{$T_{\mathrm{K}} \ [\varepsilon/k_{\mathrm{B}}]$} & \textbf{$T_0 \ [\varepsilon/k_{\mathrm{B}}]$} & \textbf{$T_{\lambda} \ [\varepsilon/k_{\mathrm{B}}]$} \\
\midrule
3DKA12 & 0.25 & 0.31 & 0.28 \\
3DKA14 & 0.60 & 0.72 & 0.65 \\
3DKA16 & 1.16 & 1.30 & 1.21 \\
3DKA18 & 1.92 & 2.22 & 2.09 \\
3DKA20 & 3.04 & 3.44 & 3.28 \\
3DIPL & 0.45 & 0.56 & 0.54 \\
2DMKA & 0.25 & 0.16 & 0.20 \\
\bottomrule
\end{tabular}

\vspace{0.4em}
\end{table}

Figure~\ref{fig2}(a) shows the maximum Lyapunov exponent $\lambda_{\mathrm{max}}$ as a function of $T$ for all the systems considered. Upon cooling, $\lambda_{\mathrm{max}}$ decreases systematically and exhibits a pronounced tendency toward a finite-temperature intercept. To characterize this behavior without imposing a specific critical exponent, we first fit the simulation data to $ \lambda_{\mathrm{max}}(T) = \lambda_0 \left(T-T_\lambda\right)^\gamma$, with $\lambda_0$, $T_\lambda$, and $\gamma$ treated as independent fitting parameters. As shown in the Supplementary Information (SI), the resulting exponents are compatible with $\gamma = 1/2$ for all models within numerical uncertainty. We therefore use the square-root form
\begin{equation}
	\lambda_{\mathrm{max}}(T) = \lambda_0\left(T - T_\lambda\right)^{1/2}
	\label{eq:sqrt_scaling}
\end{equation}
to obtain the values of $T_\lambda$ reported in Table~\ref{tab:Tk}; the corresponding extrapolations are shown as black dashed curves in Fig.~\ref{fig2}(a). The temperature range considered here lies below the onset temperature $T_A$, below which glassy dynamics becomes apparent, and above the glass transition temperature $T_{\mathrm{g}}$. It is worth noting that a square-root scaling of the largest Lyapunov exponent with the distance from the transition point has been established analytically for the stability--instability transition in Hamiltonian systems of general dimension~\cite{Hioe_1987_35_847}, so our finding of Eq.~\ref{eq:sqrt_scaling} is consistent with this universal exponent.

We find that the temperatures $T_\lambda$ obtained entirely from the chaotic dynamics closely track the independently determined Kauzmann temperatures $T_{\mathrm K}$ at which the configurational entropy extrapolates to zero (see SI). This correspondence persists across dimensionality, interaction potential, and density.

For comparison, we also consider the characteristic temperature $T_0$ associated with structural relaxation, defined as the divergence temperature obtained from the extrapolation of the structural relaxation time $\tau_\alpha$ (see SI). If the Adam--Gibbs relation between $\tau_\alpha$ and $S_c$ holds down to low temperatures, one expects $T_0$ and $T_{\mathrm K}$ to coincide. In practice, modest differences between these two temperatures are commonly found in both experiments and simulations, reflecting the substantial extent of the extrapolations involved and, in particular, the difficulty of accurately determining $S_c$ (see SI). As shown in Table~\ref{tab:Tk}, $T_{\mathrm K}$ and $T_0$ do not coincide exactly in our systems, but define closely related low-temperature scales. Remarkably, $T_\lambda$ tracks both of them across the full range of systems considered, providing an independent dynamical estimate of the characteristic temperature associated with the ideal glass transition. Consistent with this picture, in the SI we further show that the structural relaxation time is directly related to the suppression of chaotic dynamics through the approximate relation
\begin{equation}
	\ln \tau_\alpha
	\propto
 	\lambda_{\mathrm{max}}^{-2},
\end{equation}
providing an independent link between the growth of glassy relaxation times and the loss of chaos. Combining this result with the scaling relation in Eq.~\ref{eq:sqrt_scaling} leads to the relation $\ln \tau_\alpha \propto 1/(T - T_\lambda)$, which is the well-known Vogel--Fulcher--Tammann (VFT) relation that applies generally to the dynamics of glass-forming materials as a function of $T$~\cite{Supercooled_2001_410_259a,Facets_2016_69_40,Theoretical_2011_83_587a}.

The common scaling becomes particularly transparent in Fig.~\ref{fig2}(b), where the data are represented in terms of $\lambda_{\mathrm{max}}^2$ and the distance from the independently fitted Lyapunov temperature $T_\lambda$. After accounting for the non-universal amplitude $\lambda_0$, all the systems follow the aforementioned scaling $\lambda_{\mathrm{max}} \propto \left(T - T_\lambda\right)^{\gamma}$, with a common approximate exponent $\gamma = 1/2$. The amplitude $\lambda_0$ is system-dependent, whereas the square-root dependence is shared by all the systems investigated.

\begin{figure*}[htp]
	\centering
	\includegraphics[width=0.98\linewidth]{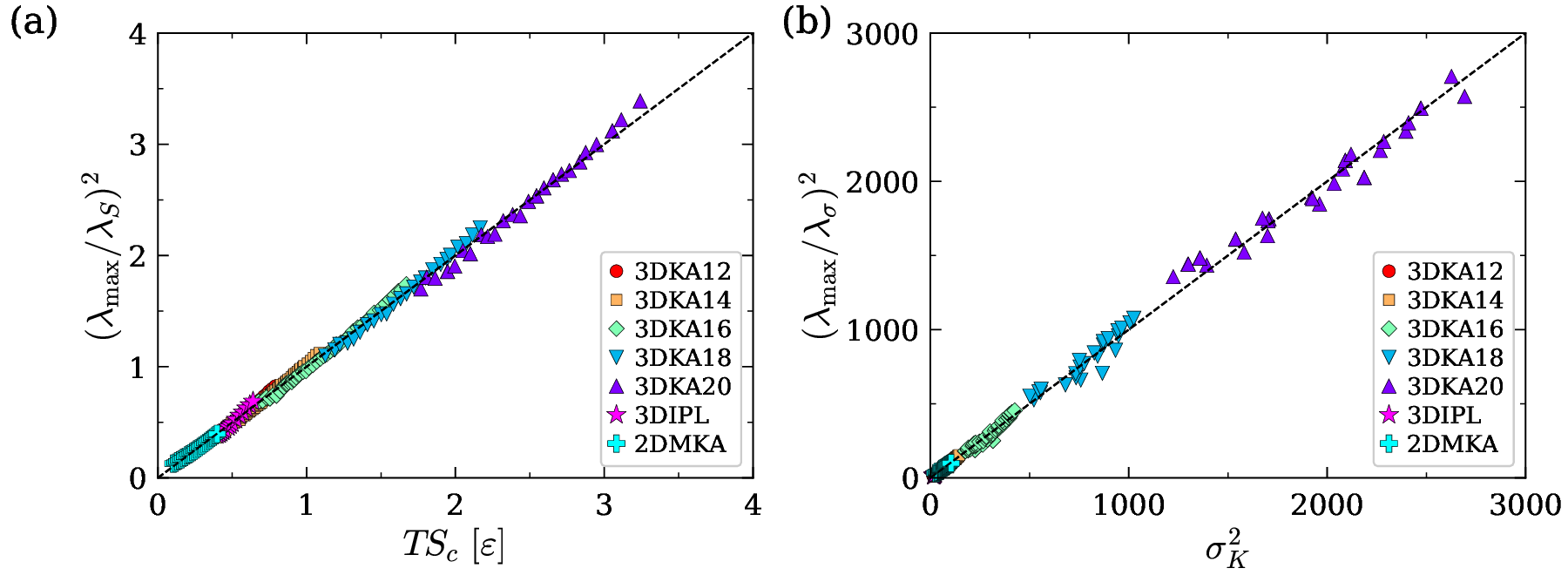}
	\caption{\textbf{Unification of dynamics, thermodynamics and potential landscape geometry.} \textbf{(a)} Reduced representation highlighting the scaling $\lambda_{\mathrm{max}}^2 \propto T S_c$ after accounting for the system-dependent prefactor $\lambda_S$. \textbf{(b)} Reduced representation highlighting the scaling $\lambda_{\mathrm{max}}^2 \propto \sigma_K^2$ after accounting for the system-dependent prefactor $\lambda_\sigma$.}
	\label{fig3}
\end{figure*}

The agreement $T_\lambda \simeq T_{\mathrm{K}}$ therefore links the extrapolated disappearance of exponential dynamical instability to the material thermodynamics. In this sense, the maximum Lyapunov exponent plays the role of a dynamical order parameter: it remains finite in the chaotic liquid and extrapolates continuously to zero as the ideal glass state is approached. The empirical exponent $1/2$ has a simple kinetic interpretation. For hard spheres at fixed density and particle diameter, the collision frequency scales with the thermal speed, $v_{\mathrm{}}\propto\sqrt{k_{\mathrm B}T/m}$ with $k_{\mathrm B}$ and $m$ being the Boltzmann constant and particle mass, and the Lyapunov exponents inherit the same $T^{1/2}$ scaling~\cite{Kinetic_2000_101_231}. By analogy, $(T-T_\lambda)$ in Eq.~\eqref{eq:sqrt_scaling} can be interpreted phenomenologically as an effective thermal-motion intensity contributing to chaotic instability. This interpretation does not identify $(T-T_\lambda)$ with the actual kinetic temperature or derive the finite intercept from hard-sphere dynamics. We next examine how this empirical scaling connects to independently determined thermodynamic and geometric quantities.

\section*{Linking Chaos to Configurational Entropy}
The configurational entropy $S_c$ quantifies the multiplicity of distinct amorphous states accessible to the system (see SI). In the thermodynamic limit, their number scales as $\mathcal{N} \sim \exp(NS_c)$, so that the progressive decrease of $S_c$ upon cooling corresponds to a loss of extensive configurational degeneracy. At the Kauzmann temperature, $S_c$ extrapolates to zero and the number of relevant amorphous states ceases to be exponential in the system size.

Figure~\ref{fig3}(a) reveals a simple quantitative relation between this thermodynamic multiplicity and chaotic dynamics. For all the systems investigated, the squared maximum Lyapunov exponent is proportional to $T S_c$,
\begin{equation}
	\lambda_{\mathrm{max}}^2 = \lambda_S^2 T S_c,
	\label{main2}
\end{equation}
with a system-dependent proportionality factor $\lambda_S$. Accordingly, after rescaling $\lambda_{\mathrm{max}}$ by $\lambda_S$, data from all models collapse onto the same linear relation shown in Fig.~\ref{fig3}(a).

Equation~\eqref{main2} establishes a direct quantitative bridge between configurational thermodynamics and chaotic dynamics. While $S_c$ measures the extensive multiplicity of amorphous states, $\lambda_{\mathrm{max}}$ measures the exponential instability of microscopic trajectories. Their relation establishes a quantitative correspondence between the loss of configurational entropy and the suppression of chaotic instability, without requiring a causal identification of one quantity with the other. Related work has also reported a robust connection between $S_c(T)$ and the Debye--Waller localization parameter $\langle u^2\rangle$ across a range of empirical silicon potentials~\cite{Configurational_2026_165_084502}. Taken together, these results suggest a broader connection between configurational entropy, particle localization, and dynamical instability: as the number of accessible amorphous states decreases, particle motion becomes increasingly localized and microscopic trajectories become progressively less chaotic.

\section*{Curvature Fluctuations Control Chaos}

The Hamiltonian dynamics of many-body systems can be reformulated in geometric terms by mapping Newtonian trajectories onto geodesics of a suitably defined curved manifold~\cite{Geometric_2000_337_237}. Within this Riemannian framework, the stability of nearby trajectories is governed by the geometry sampled along the dynamics, establishing a direct connection between chaotic motion and the local structure of the PEL.

A central result of this approach is that chaotic instability can be generated not only by locally unstable directions, but also by fluctuations of the curvature experienced along a trajectory~\cite{Riemannian_1996_54_5969}. In particular, sufficiently strong temporal curvature fluctuations can produce parametric instability and exponential separation of nearby trajectories even when the average curvature remains positive.

To investigate this geometric mechanism, we compute the instantaneous Ricci curvature $K$ along the molecular dynamics trajectories and determine both its average, $\langle K\rangle$, and its variance,
\begin{equation}
	\sigma_K^2 = \langle K^2\rangle-\langle K\rangle^2,
\end{equation}
with computational details provided in the \textit{Methods}.

Figure~\ref{fig3}(b) reveals a simple relation between chaotic instability and curvature fluctuations. For all the systems investigated in both two and three dimensions, we find
\begin{equation}
	\lambda_{\mathrm{max}}^2 = \lambda_\sigma^2 \sigma_K^2,
 	\label{main3}
\end{equation}
with a system-dependent proportionality factor $\lambda_\sigma$. This relation is consistent with the average curvature $\langle K \rangle$ remaining nearly temperature-independent near the glass transition, a behavior explicitly confirmed in the SI. After rescaling $\lambda_{\mathrm{max}}$ by $\lambda_\sigma$, the data collapse onto the common linear relation shown in Fig.~\ref{fig3}(b). Thus, the strength of chaotic instability closely tracks the magnitude of the curvature fluctuations sampled by the dynamics. 

Together with the thermodynamic relation $\lambda_{\mathrm{max}}^2 \propto T S_c$, Eq.~\eqref{main3} establishes a direct connection between configurational thermodynamics and landscape geometry. Combining the two relations gives
\begin{equation}
	T S_c \propto \lambda_{\mathrm{max}}^2 \propto \sigma_K^2,
	\label{main4}
\end{equation}
up to system-dependent prefactors. This correspondence suggests that the progressive loss of configurational multiplicity upon cooling is accompanied by a suppression of the geometric fluctuations associated with chaotic instability.

\begin{figure}[t!]
	\centering
	\includegraphics[width=0.98\linewidth]{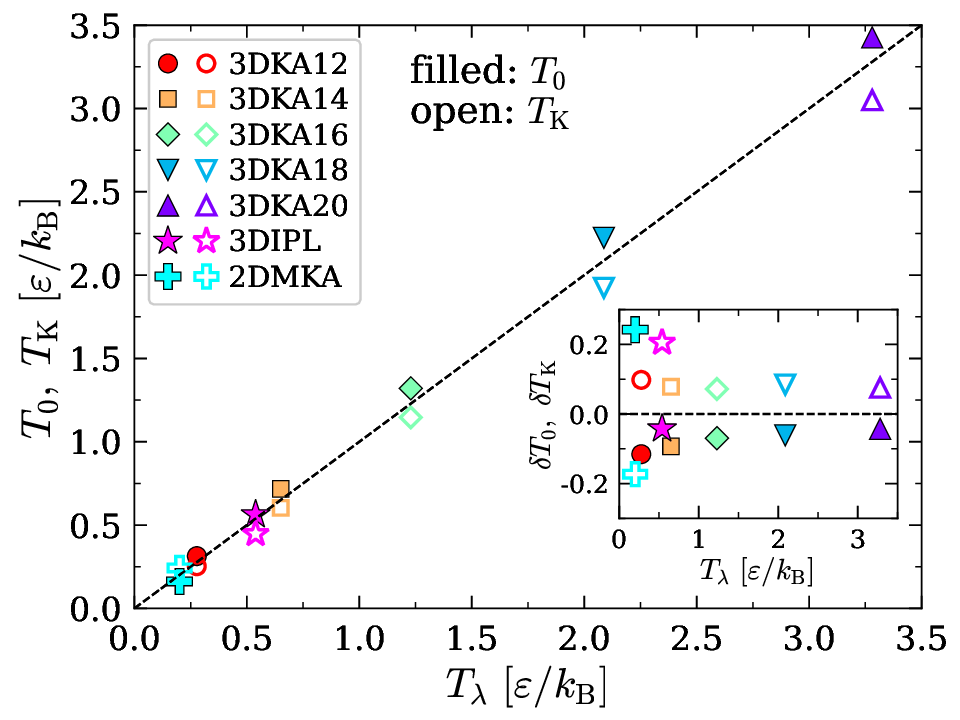}
	\caption{\textbf{Common thermodynamic and dynamical temperature scales.} Comparison of the characteristic temperature $T_\lambda$, obtained from the extrapolated vanishing of the maximum Lyapunov exponent, with the VFT temperature $T_0$ extracted from the structural relaxation dynamics (filled symbols) and the Kauzmann temperature $T_{\mathrm{K}}$ obtained by extrapolating the configurational entropy to zero (open symbols). The inset shows the relative errors $\delta T_0 \equiv (T_\lambda - T_0)/T_0$ and $\delta T_{\mathrm{K}} \equiv (T_\lambda - T_{\mathrm{K}})/T_{\mathrm{K}}$.}
	\label{fig4}
\end{figure}

The INM analysis presented in the SI yields an extrapolated temperature $T_{\mathrm{INM}}$, at which the fraction of unstable modes vanishes, with $T_{\mathrm{INM}}>T_\lambda$. This separation is consistent with chaotic dynamics persisting even in the absence of negative instantaneous modes, since temporal fluctuations of the local curvature can sustain parametric instabilities within an otherwise locally stable energy landscape~\cite{Riemannian_1996_54_5969}.

Within this picture, the approach to the ideal glass transition acquires a simple geometric interpretation. As $T$ approaches $T_\lambda$, the decrease of $\lambda_{\mathrm{max}}$ is accompanied by a corresponding suppression of curvature fluctuations, and the relation in Eq.~\eqref{main3} extrapolates both quantities toward zero. The loss of chaotic dynamics can therefore be understood geometrically as the progressive disappearance of the curvature fluctuations required to sustain a parametric instability.

\section*{A Unified View of the Ideal Glass Transition}

The results above reveal a common characteristic temperature emerging independently from thermodynamics, structural relaxation, and chaotic dynamics. As summarized in Fig.~\ref{fig4}, the Lyapunov temperature $T_\lambda$ closely tracks both the Kauzmann temperature $T_{\mathrm{K}}$ and the VFT temperature $T_0$ across all the systems and densities considered, $T_\lambda \simeq T_{\mathrm{K}} \simeq T_0$. The loss of chaotic instability is therefore tied to the same characteristic temperature scale associated with the extrapolated vanishing of configurational entropy and the dramatic slowing down of structural relaxation.

At the level of the underlying observables, the thermodynamic, dynamical, and geometric descriptions are connected by the chain of relations in Eq.~\eqref{main4}. Thus, the progressive loss of configurational multiplicity upon cooling is accompanied by a suppression of chaotic instability and of the curvature fluctuations capable of sustaining parametric instability. The distinct INM and Lyapunov extrapolation temperatures are consistent with curvature fluctuations sustaining chaos even in the absence of instantaneous unstable modes.

Taken together, these results provide a unified picture of the approach to the ideal glass state. Within the conventional extrapolated Kauzmann scenario, the ideal-glass endpoint emerges not only as an entropy crisis, but also as the endpoint of chaotic exploration of the potential energy landscape, where thermodynamic degeneracy, dynamical instability, and geometric fluctuations converge toward a common endpoint.

\section*{Methods}

\subsection*{Simulated Systems}

We performed extensive molecular dynamics simulations of three canonical glass-forming models spanning different spatial dimensions and interaction types: (i) a two-dimensional modified Kob--Andersen (2DMKA) mixture with 65:35 composition; (ii) the standard three-dimensional Kob--Andersen (3DKA) mixture with 80:20 composition at five number densities $\rho = 1.2, 1.4, 1.6, 1.8$, and $2.0~\sigma^{-3}$, labeled as 3DKA12, 3DKA14, 3DKA16, 3DKA18, and 3DKA20, respectively; and (iii) a three-dimensional binary inverse-power-law (3DIPL) system with purely repulsive interactions. The maximum Lyapunov exponent was computed for a system size of $N=1000$ in all systems. The 2DMKA and 3DIPL systems were simulated at $\rho = 1.2~\sigma^{-2}$ and $1.2~\sigma^{-3}$, respectively. Each state point was carefully equilibrated at the target temperature prior to production runs. All quantities are reported in standard reduced units. More details are given in the SI.

\subsection*{Computation of the Maximum Lyapunov Exponent}

For each state point, the system was first equilibrated at temperature $T$ using a thermostatted simulation. The maximum Lyapunov exponent $\lambda_{\mathrm{max}}$ was then evaluated along the corresponding production trajectories. The equations of motion in phase space are represented by $\dot{\mathbf{X}} = \mathbf{F}(\mathbf{X})$. The tangent space vector $\bm{\delta}(t) = [\delta \mathbf{r}(t), \delta \mathbf{p}(t)]$ naturally decomposes into coordinate- and momentum-space components, which describe the projection of the locally unstable direction onto the two sectors of phase space. It evolves according to the linearized equations:
\begin{equation}
	\dot{\bm{\delta}}(t) = \mathbf{J}[\mathbf{X}(t)] \bm{\delta}(t),
\end{equation}
where the Jacobian is
\begin{equation}
	\mathbf{J}(\mathbf{X})=\frac{\partial\mathbf{F}}{\partial\mathbf{X}}
	=\begin{pmatrix}
	\mathbf{0} & \mathbf{M}^{-1} \\
	-\mathbf{H}_V & \mathbf{0}
	\end{pmatrix}.
\end{equation}
Here, $\mathbf{H}_V$ is the Hessian matrix and $\mathbf{M}$ is the mass matrix. We calculated $\lambda_{\mathrm{max}}$ in Eq.~\eqref{Eq_Lmax} using the Benettin algorithm, with a periodic renormalization of the tangent vector. Taking advantage of modern GPU acceleration and the differentiable physics framework JAX-MD~\cite{JAXMD_2020_33_11479a}, we computed the Jacobian and Hessian matrices on the fly, allowing us to perform long-time tangent space integration for relatively large systems (e.g., $N = 1000$).

\subsection*{Determination of the Kauzmann Temperature}

Simulation data of the configurational entropy $S_c$ and the structural relaxation time $\tau_\alpha$ for the 3DKA, 2DMKA, and 3DIPL systems were taken from previous studies~\cite{Understanding_2024_128_10999, glassy_2024_161_34504, Scaling_2004_16_L489}. The Kauzmann temperature $T_{\mathrm K}$ was obtained by fitting the simulation data for the configurational entropy $S_c$ to the equation $T S_c = K_T (T/T_{\mathrm K} - 1)$, where $K_T$ defines a thermodynamic fragility parameter. The structural relaxation time $\tau_\alpha$ was fitted to the VFT form, $\tau_\alpha = \tau_0 \exp[D T_0 / (T - T_0)]$, to extract the VFT temperature $T_0$. For the 3DIPL system, the inverse self-diffusion coefficient $D_s^{-1}$ was used in place of $\tau_\alpha$. The values of $K_T$, $\tau_0$, and $D$ are provided in the SI.

\subsection*{Determination of the PEL Geometry}

The instantaneous Ricci curvature $K(t)$ is evaluated as the trace of the potential-energy Hessian per particle per spatial dimension. From production trajectories, we computed the mean $\langle K \rangle$ and the variance $\sigma_K^2 = \langle K^2\rangle - \langle K\rangle^2$. The mean curvature remains positive across all temperatures studied, consistent with the interpretation that the suppression of chaotic instability is associated with diminishing temporal curvature fluctuations rather than with a change in the sign of the average curvature. More details can be found in the SI.

\section*{Data availability}

The data used to generate the figures presented in this manuscript are available from the Open Science Framework: \url{https://osf.io/8s42m/}.

\section*{Code availability}

The code and scripts used to compute the Lyapunov spectrum and curvature observables are available on GitHub: \url{https://github.com/zlw-sci/Chaos_Glass}.

\begin{acknowledgments}
This work was supported by the National Natural Science Foundation of China (Nos. 22573102 and 52293471) and the Jilin Provincial Science and Technology Development Program (Nos. SKL202602006JC and SKL202502004JC). Z.-Y.S. is grateful for the essential support of Jilin Provincial International Cooperation Key Laboratory for Polymer Processing Physics. M.B. acknowledges the support of the Foreign Young Scholars Research Fund Project (Grant No. 22Z033100604) and the sponsorship from the Yangyang Development Fund.
\end{acknowledgments}

\section*{Author contributions}

W.-S.X., M.B., Z.-Y.S., and J.F.D. designed the project. Z.-L.W. implemented the method to compute the Lyapunov spectrum. All authors analyzed the data, discussed the results, and contributed to the writing of the manuscript.

\section*{Competing interests}

The authors declare no competing interests.

\clearpage
\onecolumngrid

\setcounter{equation}{0}
\setcounter{figure}{0}
\setcounter{table}{0}
\setcounter{section}{0}
\renewcommand{\theequation}{S\arabic{equation}}
\renewcommand{\thefigure}{S\arabic{figure}}
\renewcommand{\thetable}{S\arabic{table}}
\renewcommand{\thesection}{S\arabic{section}}

\begin{center}
{\large\textbf{Supplementary Information for\\ ``The End of Chaos at the Ideal Glass Transition''}}
\end{center}
\vspace{1em}

\section{Models and Simulation Details}
\label{sec:models}

We consider a 2D modified Kob--Andersen mixture (2DMKA), the canonical 3D Kob--Andersen mixture (3DKA) at several densities, and a 3D binary inverse-power-law mixture (3DIPL). Simulation data for the configurational entropy and relaxation dynamics of these systems can be found in Refs.~\cite{Influence_2024_128_9889, Understanding_2024_128_10999, glassy_2024_161_34504, Scaling_2004_16_L489}.

The 3DKA interaction is governed by the truncated and shifted Lennard--Jones (LJ) potential
\begin{equation}
U_{\alpha\beta}^{\mathrm{KA}}(r)=
\begin{cases}
4\varepsilon_{\alpha\beta}
\left(x^{-12}-x^{-6}-x_c^{-12}+x_c^{-6}\right),&x<x_c,\\
0,&x\ge x_c,
\end{cases}
\end{equation}
where $x=r/\sigma_{\alpha\beta}$ and the cutoff distance is set to $x_c=2.5$ for all models. The composition is 80:20. More details can be found in Ref.~\cite{Influence_2024_128_9889}.

For the 2DMKA system, the composition is 65:35, and the interaction potential is smoothed to be force-continuous at the cutoff:
\begin{equation}
U_{\alpha\beta}^{\mathrm{MKA}}(r)=
\begin{cases}
4\varepsilon_{\alpha\beta}
\left(x^{-12}-x^{-6}+c_0+c_2x^2\right),&x<x_c,\\
0,&x\ge x_c,
\end{cases}
\end{equation}
where $c_0=-x_c^{-12}+x_c^{-6}-c_2x_c^2$ and $c_2=6x_c^{-14}-3x_c^{-8}$. Further details can be found in Ref.~\cite{glassy_2024_161_34504}.

The interaction of the 3DIPL system is given by
\begin{equation}
U_{\alpha\beta}^{\mathrm{IPL}}(r)=
\begin{cases}
4\varepsilon_{\alpha\beta}
\left(x^{-12}+c_0+c_2x^2+c_4x^4\right),&x<x_c,\\
0,&x\ge x_c,
\end{cases}
\end{equation}
where $c_0$, $c_2$, and $c_4$ ensure that the potential and its first two derivatives vanish at the cutoff distance. The composition is 80:20. %Additional details are available in Ref.~\citenum{Scaling_2004_16_L489}.

For each system, all particles have the same mass $m$. We set $\sigma_{AA}=\varepsilon_{AA}=m=k_{\mathrm B}=1$ in our simulations. We report all quantities in standard reduced units, with length, energy, and time expressed in units of \(\sigma\equiv\sigma_{AA}\), \(\varepsilon\equiv\varepsilon_{AA}\), and \(\tau\equiv\sigma_{AA}\sqrt{m/\varepsilon_{AA}}\), respectively. For all the binary mixtures, the interaction parameters are $\sigma_{AA}=1 \ \sigma$, $\sigma_{AB}=0.8 \ \sigma$, $\sigma_{BB}=0.88 \ \sigma$, $\varepsilon_{AA}=1 \ \varepsilon$, $\varepsilon_{AB}=1.5 \ \varepsilon$, and $\varepsilon_{BB}=0.5 \ \varepsilon$. Here, the subscripts $A$ and $B$ denote particle species. For the 3DKA system, five number densities are investigated: $\rho = N/V = 1.2, 1.4, 1.6, 1.8$, and $2.0 \ \sigma^{-3}$, labeled as 3DKA12, 3DKA14, 3DKA16, 3DKA18, and 3DKA20, respectively. Here, $V$ is the volume or area of the simulation box. The 2DMKA and 3DIPL systems are simulated at $\rho = 1.2 \ \sigma^{-2}$ and $\rho = 1.2 \ \sigma^{-3}$, respectively. The total number of particles is $N = 1000$ for all the systems in our calculation of the maximum Lyapunov exponent, although we also consider a smaller system of $N=200$ for illustration purposes of the Lyapunov spectrum. 

The system at each state point was equilibrated at the target temperature before the calculations of properties were performed. For the 3DIPL system, canonical stochastic velocity rescaling at intervals of $100 \,\Delta t$ was applied for $10^6$ time steps, followed by a production trajectory of $2 \times 10^6$ time steps in the $NVE$ ensemble, where the time step is $\Delta t = 0.002 \ \tau$. Periodic boundary conditions were applied along all directions in the simulations.

\section{Lyapunov Spectrum and Dynamical Entropy}
\label{sec:lyapunov-components}

To characterize the sensitivity of the dynamics to infinitesimal perturbations, we consider the evolution of nearby trajectories in phase space. For a system of particles described by positions $\mathbf r$ and momenta $\mathbf p$, an infinitesimal deviation between two trajectories generally contains both a coordinate and a momentum component. The corresponding tangent vector can therefore be written as
\begin{equation}
\bm{\delta}(t)=\bigl(\delta\mathbf r(t),\delta\mathbf p(t)\bigr).
\end{equation}
The coordinate component $\delta\mathbf r$ describes the separation of particle positions, while $\delta\mathbf p$ describes the corresponding separation in momentum space. Their coupled evolution determines the growth of the full phase-space perturbation and hence the Lyapunov exponents.

It is important to emphasize that the coordinate and momentum contributions do not define independently additive Lyapunov exponents. The exponents characterize the asymptotic growth rates of perturbations in the full tangent space. In the reduced coordinates and momenta employed here, the corresponding Euclidean norm is
\begin{equation}
|\bm{\delta}|^2=
|\delta\mathbf r|^2+|\delta\mathbf p|^2.
\end{equation}

\begin{figure}[htp!]
\centering
\includegraphics[width=0.96\linewidth]{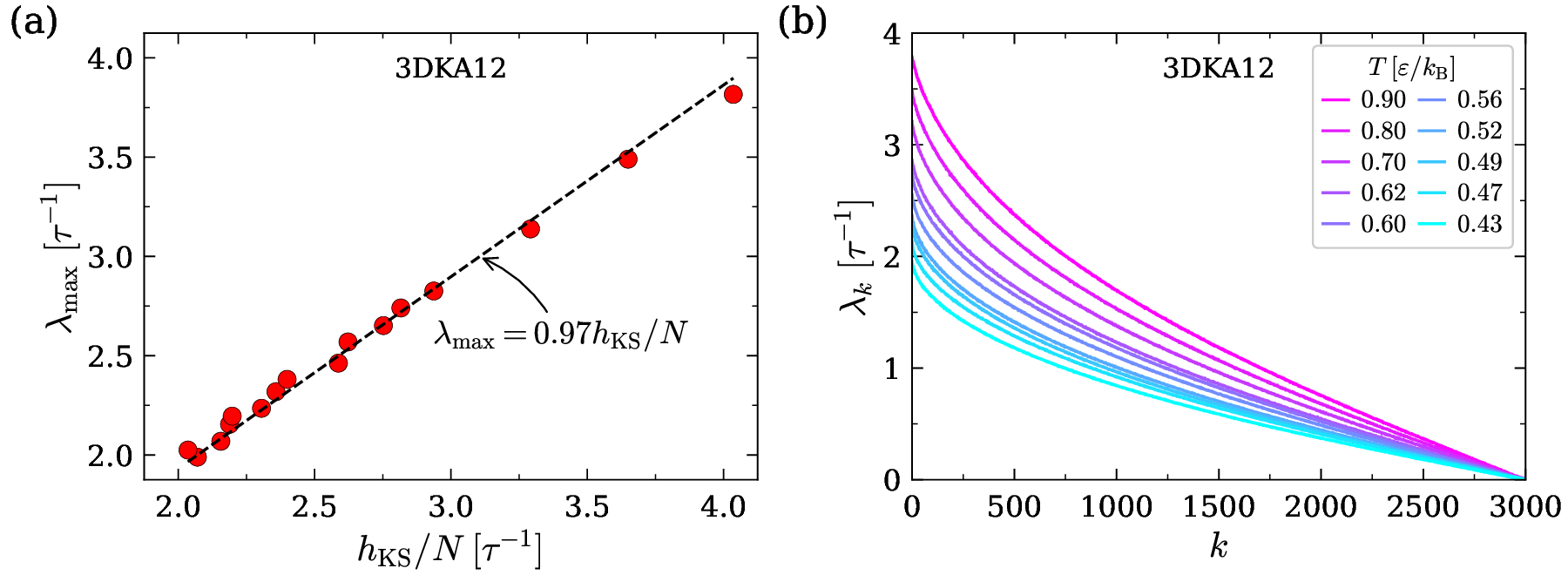}
\caption{\textbf{Relation between the maximum Lyapunov exponent and the dynamical entropy rate.} \textbf{(a)} $\lambda_{\mathrm{max}}$ versus the Kolmogorov--Sinai entropy rate $h_\mathrm{KS}$ normalized by the particle number $N$ for the 3DKA12 system. The dashed line is a linear fit to the simulation data, as indicated in the figure. \textbf{(b)} Lyapunov spectrum $\{\lambda_k\}$ for the 3DKA12 system over a range of $T$. Only the positive part of the spectrum is considered since this information is sufficient to determine $h_\mathrm{KS}$.}
\label{fig:hk}
\end{figure}

The Lyapunov spectrum characterizes the growth and decay rates of infinitesimal perturbations in phase space. For Hamiltonian dynamics, the nonzero exponents occur in pairs of opposite sign, corresponding to expanding and contracting directions, while zero exponents are associated with neutral directions. Accordingly, the sum over the full spectrum vanishes, as required by conservation of phase-space volume. The positive part of the spectrum defines
\begin{equation}
h_\mathrm{KS}=\sum_{\lambda_i>0}\lambda_i.
\label{eq:hk}
\end{equation}
Under the conditions for Pesin's identity, this quantity coincides with the Kolmogorov--Sinai entropy rate $h_\mathrm{KS}$~\cite{Extensivity_2017_95_22102}. While $\lambda_{\mathrm{max}}$ captures the fastest exponential divergence of nearby trajectories, $h_\mathrm{KS}$ incorporates all unstable directions and therefore provides a measure of the total dynamical information-production rate. This dynamical entropy should not be confused with the thermodynamic configurational entropy $S_c$.

In Fig.~\ref{fig:hk}(a), we show that $\lambda_{\mathrm{max}}$ is, to a good approximation, proportional to $h_\mathrm{KS}/N$ over the temperature range considered. Because computing the full Lyapunov spectrum is numerically demanding, we restrict this analysis to the positive branch of the spectrum and to the 3DKA12 system, shown in Fig.~\ref{fig:hk}(b). The progressive suppression of chaotic dynamics upon cooling is therefore visible not only in the largest Lyapunov exponent, but also in the collective contribution of all unstable directions, providing an independent complement to the behavior of $\lambda_{\mathrm{max}}$ discussed in the main text.

\section{Power-Law Analysis of the Maximum Lyapunov Exponent}
\label{sec:gamma}

\begin{figure}[htp!]
\centering
\includegraphics[width=0.62\linewidth]{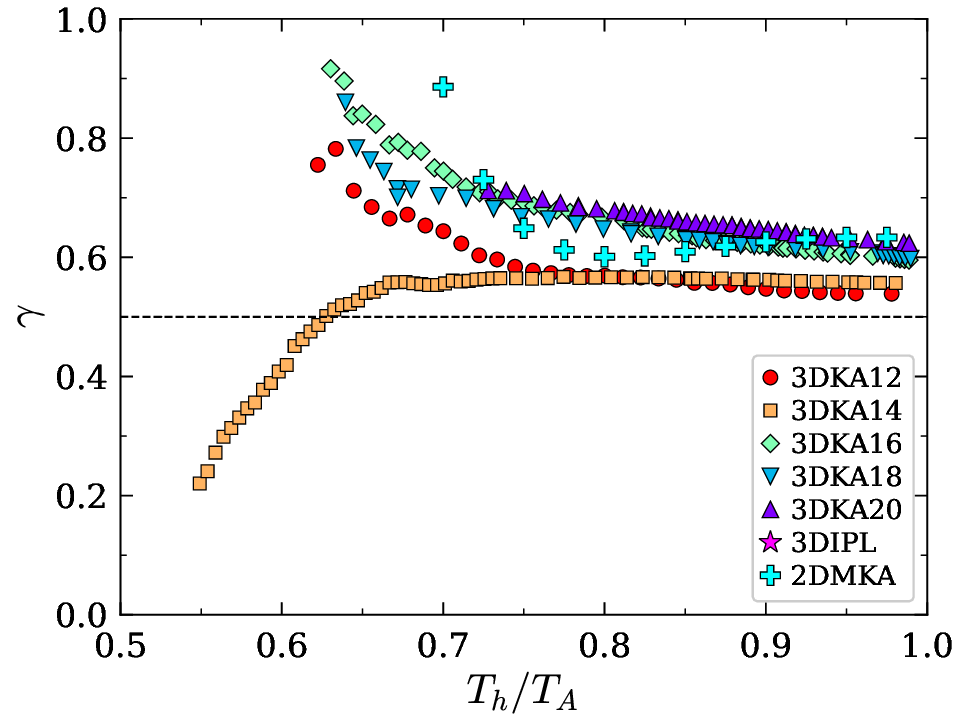}
\caption{\textbf{Power-law scaling of the maximum Lyapunov exponent.} \label{fig:gamma}The free power-law exponent $\gamma$ in Eq.~\eqref{eq:freegamma} as a function of $T_h/T_A$, where $T_h$ is the upper boundary of the fitting interval and $T_A$ is the onset temperature of glass formation, for all the systems studied. The horizontal dashed line marks the approximate square-root behavior used in the main text, $\gamma=1/2$.}
\end{figure}

\begin{table}[htbp!]
	\centering
	\caption{Summary of the fitted parameters. $T_A$ is the onset temperature of glass formation at which non-Arrhenius dynamics starts to emerge. $\lambda_0$, $\lambda_S$, and $\lambda_{\sigma}$ are the prefactors in Eqs.~(\ref{eq:sqrt_scaling}), (\ref{main2}), and (\ref{main3}) of the main text. $K_T$ is the parameter in Eq.~\eqref{eq:scfit}. $\tau_o$ is the prefactor in the Adam--Gibbs relation. $\tau_0$ and $D$ are the VFT parameters in Eq.~\eqref{eq:vft}.}
	\label{Table_Fits}
	\begin{ruledtabular}
	\begin{tabular}{lcccccccc}
	System & $T_A$ & $\lambda_0$ & $\lambda_S$ & $\lambda_{\sigma}$ & $K_T$ & $\tau_o$ & $\tau_0$ & $D$ \\
	\colrule
	3DKA12 & 0.90 & 4.84 & 4.53 & 0.55 & 0.27 & $2.39 \times 10^{-1}$ & $4.79 \times 10^{-1}$ & 3.75 \\
	3DKA14 & 2.04 & 4.89 & 4.50 & 0.38 & 0.66 & $1.45 \times 10^{-1}$ & $4.51 \times 10^{-1}$ & 3.08 \\
	3DKA16 & 3.57 & 4.94 & 4.49 & 0.28 & 1.28 & $8.05 \times 10^{-2}$ & $2.95 \times 10^{-1}$ & 3.30 \\
	3DKA18 & 5.88 & 5.00 & 4.50 & 0.21 & 2.12 & $2.63 \times 10^{-2}$ & $1.82 \times 10^{-1}$ & 3.60 \\
	3DKA20 & 8.93 & 5.05 & 4.55 & 0.16 & 3.37 & $1.64 \times 10^{-2}$ & $9.49 \times 10^{-2}$ & 4.17 \\
	3DIPL & 1.00 & 4.30 & 3.45 & 0.54 & 0.54 & $2.41$ & $18.1$ & 2.77 \\
	2DMKA & 0.80 & 3.87 & 5.16 & 0.33 & 0.15 & $1.27$ & $0.36$ & 14.8 \\
	\end{tabular}
	\end{ruledtabular}
\end{table}

To extract the temperature dependence without constraining the power-law exponent, we fit the simulation data to
\begin{equation}
\lambda_{\mathrm{max}} \propto \left(T-T_\lambda\right)^\gamma.
\label{eq:freegamma}
\end{equation}
Figure~\ref{fig:gamma} shows the fitted exponent as the upper boundary $T_h$ of the fitting interval is varied. Although the fitted exponent depends somewhat on the fitting window, it remains close to \(1/2\) over a broad low-temperature range. We therefore use $\gamma=1/2$ as a common empirical representation,
\begin{equation}
\lambda_{\mathrm{max}}^2=
\lambda_0^2\left(T-T_\lambda\right),
\label{eq:fixedgamma}
\end{equation}
which defines the characteristic temperature $T_\lambda$ quoted in the main text. The fitted values of $\lambda_0$ are listed in Table~\ref{Table_Fits}.

The square-root approximation is motivated by hard-sphere collision dynamics: at fixed density and particle diameter, both the collision frequency and the Lyapunov exponents scale as $T^{1/2}$~\cite{Kinetic_2000_101_231}. For the present glass-forming liquids, the shifted scale $T-T_\lambda$ provides a phenomenological measure of the effective thermal-motion intensity contributing to chaotic instability. The finite offset is determined empirically, so it does not imply that thermal motion itself ceases at $T_\lambda$.

\section{Determination of the Kauzmann and VFT Temperatures}
\label{sec:entropy}

The dynamical and thermodynamic data used in our analysis were taken from previous studies. The density-dependent 3DKA data were reported by Yuan et al.~\cite{Influence_2024_128_9889,Understanding_2024_128_10999}, the 2DMKA data by Nath and Sengupta~\cite{glassy_2024_161_34504}, and the 3DIPL data by De Michele et al.~\cite{Scaling_2004_16_L489}. 

\begin{figure}[tb]
\centering
\includegraphics[width=0.96\linewidth]{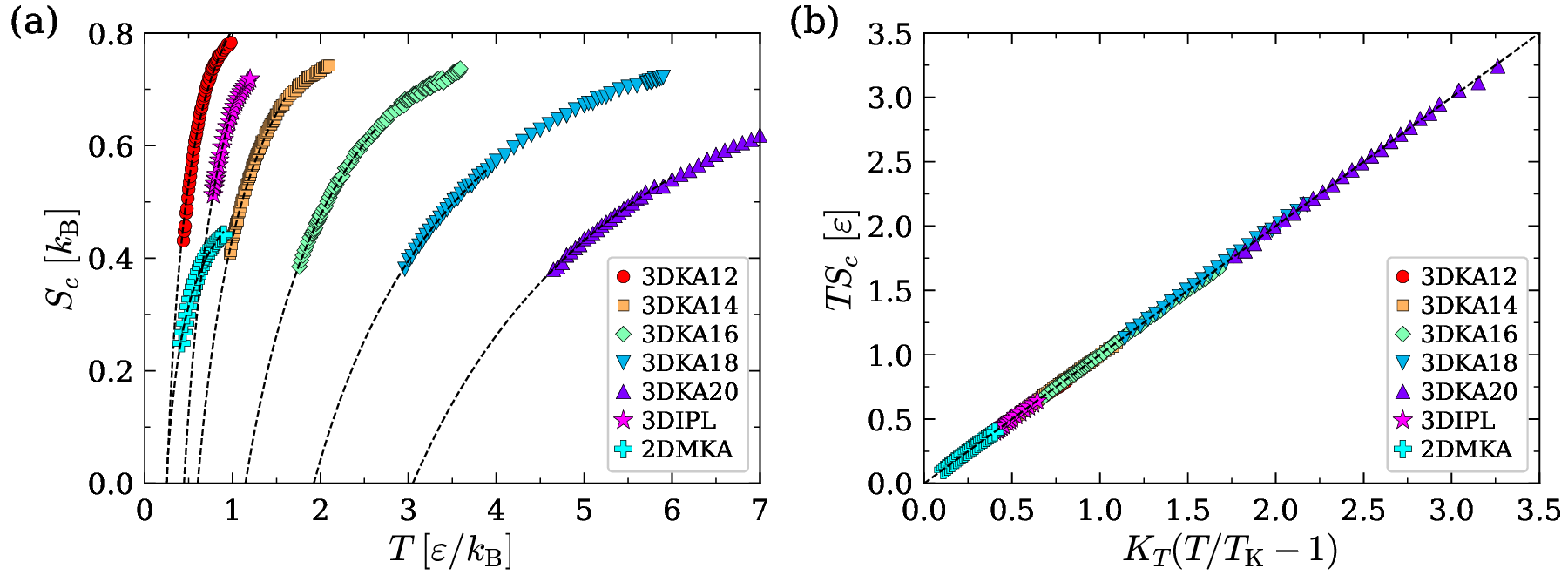}
\caption{\label{fig:sc}\textbf{Determination of the Kauzmann temperature from the configurational entropy.} \textbf{(a)} Temperature dependence of $S_c$ for all the systems studied. Dashed curves are fits to Eq.~\eqref{eq:scfit}. \textbf{(b)} Reduced representation of the same data, showing $T S_c$ against $K_T(T/T_{\mathrm{K}}-1)$. The dashed line denotes equality.}
\end{figure}

The configurational entropy was evaluated as
\begin{equation}
S_c=S_{\mathrm{tot}}-S_{\mathrm{vib}},
\end{equation}
where $S_{\mathrm{tot}}$ and $S_{\mathrm{vib}}$ are the total and vibrational entropies, respectively. The Kauzmann temperature was obtained by extrapolating the configurational-entropy data using the following equation:
\begin{equation}
T S_c=K_T\left(\frac{T}{T_{\mathrm{K}}}-1\right),
\label{eq:scfit}
\end{equation}
where $K_T$ defines a thermodynamic fragility parameter. Figure~\ref{fig:sc} shows the individual entropy curves and the corresponding reduced representation. These $T_{\mathrm{K}}$ values are reported in the main text. The fitted values of $K_T$ are listed in Table~\ref{Table_Fits}.

\begin{figure}[tb]
\centering
\includegraphics[width=0.96\linewidth]{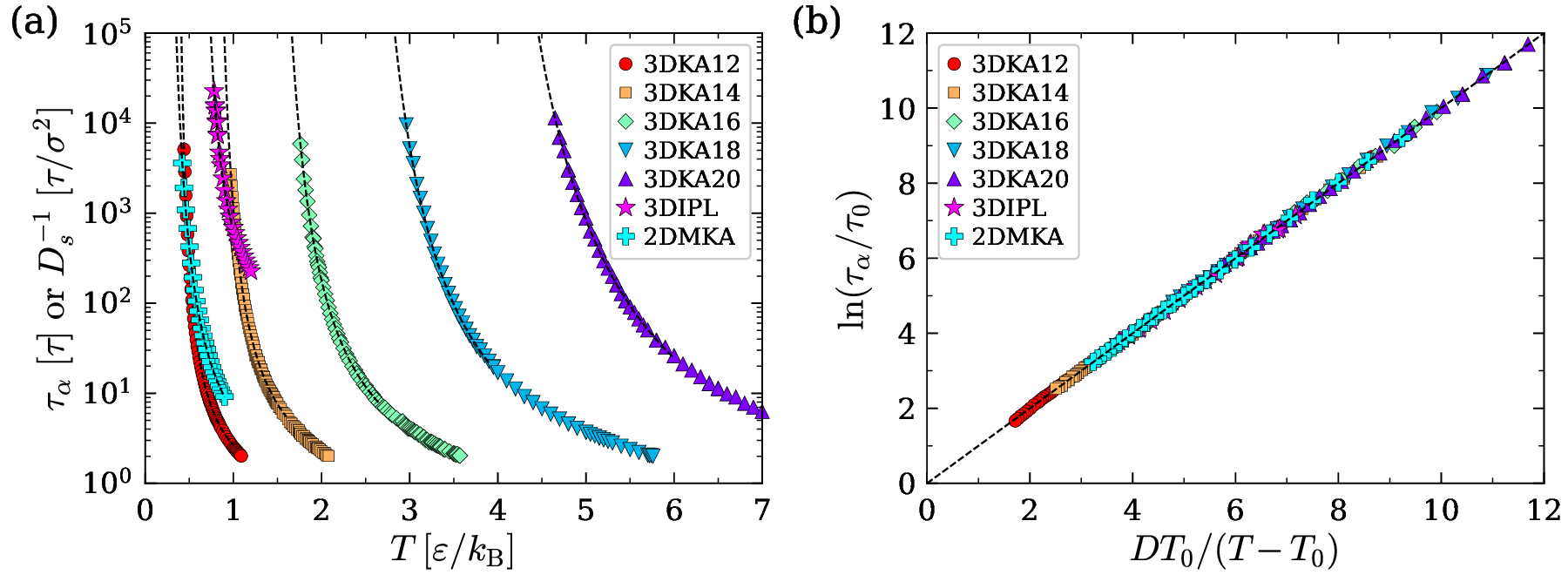}
\caption{\label{fig:vft}\textbf{Determination of the VFT temperature from the structural relaxation time.} \textbf{(a)} Temperature dependence of $\tau_\alpha$ for the 3DKA and 2DMKA systems and of $D_s^{-1}$ for the 3DIPL system. Dashed curves are fits to Eq.~\eqref{eq:vft}. \textbf{(b)} VFT representation of the same data. The dashed line denotes equality.}
\end{figure}

The VFT temperature $T_0$ is determined from the $T$ dependence of dynamical timescales. For the 3DKA and 2DMKA systems, the characteristic dynamical timescale is the structural relaxation time $\tau_\alpha$. For the 3DIPL system, the inverse self-diffusion coefficient $D_s^{-1}$ is used instead; in the following equations, $\tau_\alpha$ and $\tau_0$ are then understood as $D_s^{-1}$ and its VFT prefactor, respectively. The temperature dependence of the corresponding dynamical timescale was fitted to the Vogel--Fulcher--Tammann (VFT) form
\begin{equation}
\tau_\alpha=\tau_0
\exp\left(\frac{D T_0}{T-T_0}\right),
\label{eq:vft}
\end{equation}
where $\tau_0$ is a prefactor, $D$ is the fragility parameter, and $T_0$ is the extrapolated divergence temperature. As shown in Fig.~\ref{fig:vft}(b), the transformed data follow the linear representation of Eq.~\eqref{eq:vft} over the fitted range. The fitted results for $T_0$ are provided in the main text, and $\tau_0$ and $D$ are summarized in Table~\ref{Table_Fits}.

\section{Relation between the Maximum Lyapunov Exponent and Structural Relaxation Time}
\label{sec:product}

\begin{figure}[tb]
\centering
\includegraphics[width=0.62\linewidth]{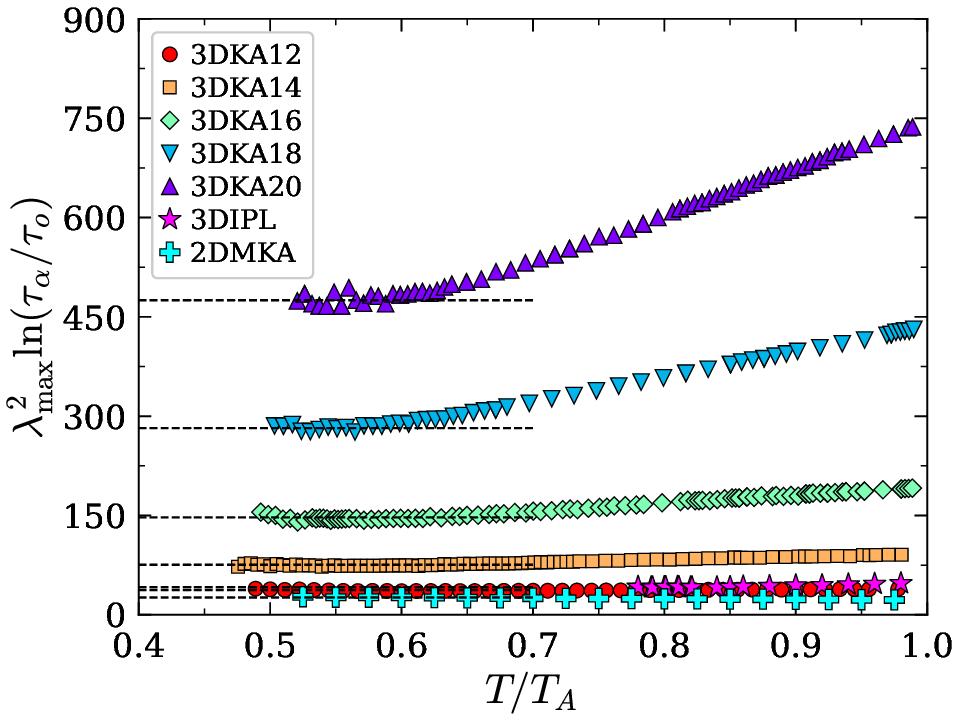}
\caption{\textbf{Consistency between the maximum Lyapunov exponent and structural relaxation time.} \label{fig:product}The product
$\lambda_{\mathrm{max}}^2\ln(\tau_\alpha/\tau_o)$ as a function of $T/T_A$ for all the systems studied. Horizontal dashed lines indicate the approximate plateaus in the low-temperature regime. For the 3DIPL system, $\tau_\alpha$ is replaced by $D_s^{-1}$.}
\end{figure}

Combining the scaling $\lambda_{\mathrm{max}}^2\propto T S_c$ and the Adam--Gibbs relation, $\ln(\tau_\alpha/\tau_o)\propto 1/(T S_c)$, where $\tau_o$ is the prefactor (listed in Table~\ref{Table_Fits}), leads to the following relation:
\begin{equation}
\lambda_{\mathrm{max}}^2
\ln\left(\frac{\tau_\alpha}{\tau_o}\right)
= \mathrm{constant}.
\label{eq:product1}
\end{equation}
Figure~\ref{fig:product} shows that an approximate plateau is observed in the lower-temperature part of the common fitting window, while deviations develop at higher temperatures. Note that the Adam--Gibbs relation does not hold exactly in the 2DMKA system~\cite{glassy_2024_161_34504}. Nevertheless, we adopt this relation here as a reasonable approximation.

Alternatively, combining the Lyapunov scaling in Eq.~\eqref{eq:fixedgamma} with the VFT relation yields
\begin{equation}
\lambda_{\mathrm{max}}^2
\ln\left(\frac{\tau_\alpha}{\tau_0}\right)
=\lambda_0^2\,D\, T_0\,
\frac{T-T_\lambda}{T-T_0}.
\label{eq:product}
\end{equation}
When \(T_\lambda\) and \(T_0\) are close, as found here (see Table~\ref{tab:Tk} in the main text), the product on the left-hand side is approximately temperature independent. 

\section{Curvature Fluctuations of the Potential Energy Landscape}
\label{sec:curvature}

\begin{figure}[tb]
\centering
\includegraphics[width=0.62\linewidth]{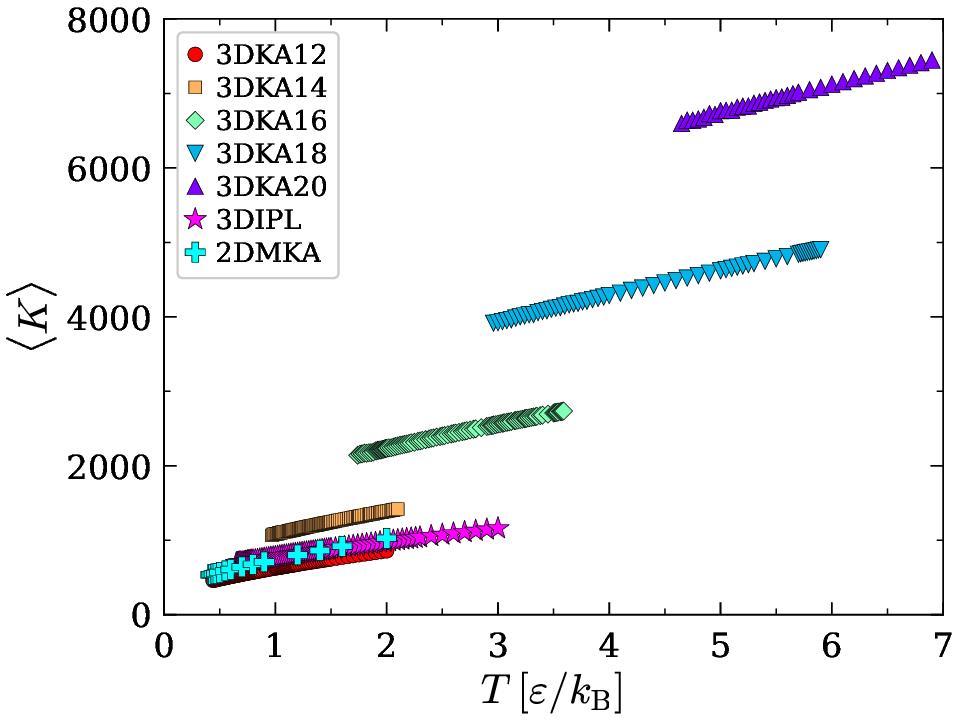}
\caption{\label{fig:kmean}\textbf{Mean curvature of the PEL.} Temperature dependence of the Ricci curvature $\langle K\rangle$ per configurational degree of freedom for all the systems studied.}
\end{figure}

The instantaneous Ricci curvature is the trace of the potential-energy Hessian, $K_R(t)=\operatorname{Tr}\mathbf H_V[\mathbf r(t)]$. The quantity evaluated in the simulations is the curvature per configurational degree of freedom,
\begin{equation}
K(t)=\frac{K_R(t)}{dN}
=\frac{\operatorname{Tr}\mathbf H_V[\mathbf r(t)]}{dN},
\label{eq:kcurve}
\end{equation}
where $d$ is the spatial dimension and $N$ is the number of particles. For the pairwise potentials used here, the same quantity is evaluated directly as
\begin{equation}
K(t)=\frac{2}{dN}\sum_{i<j}\left[
v_{ij}''(r_{ij})+\frac{d-1}{r_{ij}}v_{ij}'(r_{ij})\right].
\label{eq:kcurve-pair}
\end{equation}
The factor of two accounts for the two diagonal Hessian blocks associated with each interacting pair. From the sampled production frames, we calculated
\begin{equation}
\langle K\rangle=\frac{1}{M}\sum_{m=1}^{M}K(t_m)
\label{eq:kmean}
\end{equation}
and the temporal variance of this spatially averaged curvature,
\begin{equation}
\sigma_K^2=\left\langle K^2\right\rangle-\left\langle K\right\rangle^2.
\label{eq:sigmak}
\end{equation}
Figure~\ref{fig:kmean} shows that the mean curvature remains positive over the investigated temperature range. The suppression of chaotic instability is therefore associated with a reduction of temporal curvature fluctuations rather than with a sign change or vanishing of the mean curvature.

\section{Instantaneous Normal Modes}
\label{sec:inm}

\begin{figure}[tb]
\centering
\includegraphics[width=0.96\linewidth]{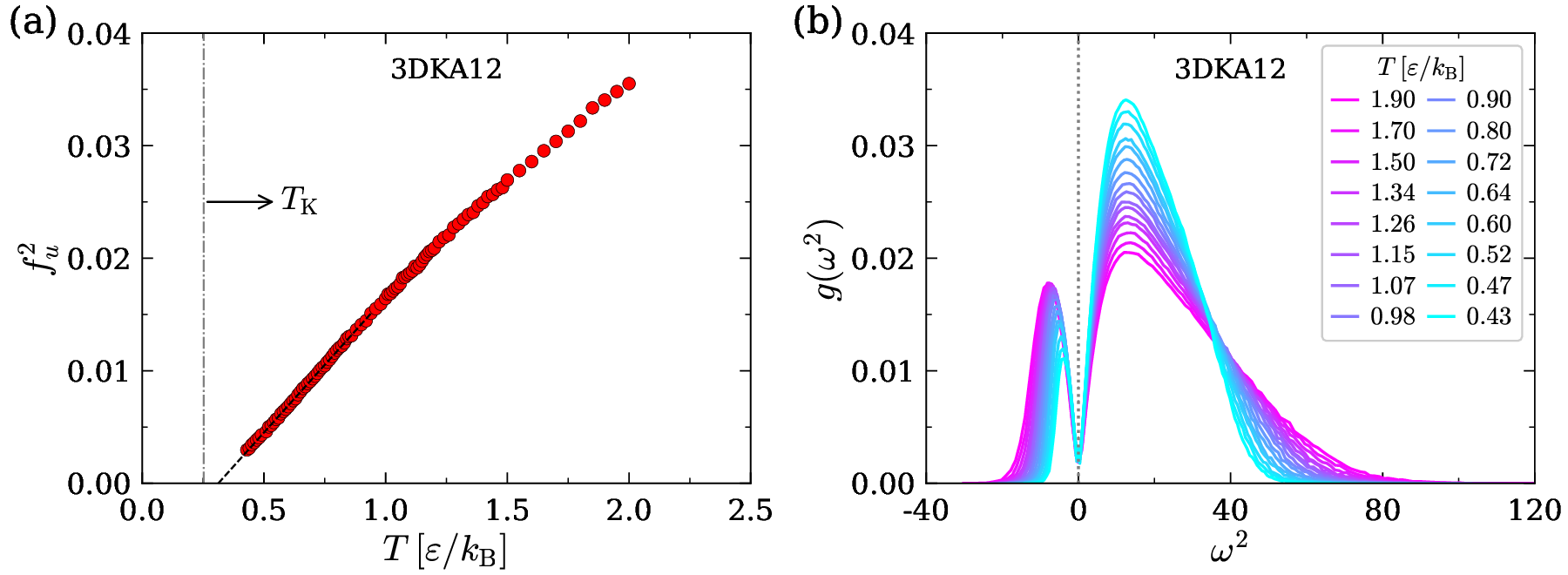}
\caption{\label{fig:fu_inm}\textbf{Instantaneous normal modes.} \textbf{(a)} Temperature dependence of $f_u^2$ for the 3DKA12 system. The black dashed line denotes the low-temperature linear extrapolation used to determine $T_{\mathrm{INM}}$. The vertical grey line indicates the Kauzmann temperature $T_{\mathrm{K}}$. The low-temperature extrapolation of the unstable-mode fraction vanishes at $T_{\mathrm{INM}}$ before reaching $T_{\mathrm{K}}$ upon cooling (i.e., $T_{\mathrm{INM}} > T_{\mathrm{K}}$). \textbf{(b)} Eigenvalue density $g(\omega^2)$ for the 3DKA12 system over a range of $T$. The negative $\omega^2$ sector corresponds to unstable modes.}
\end{figure}

At each sampled configuration, the eigenvalues $\omega_\alpha^2$ of the instantaneous Hessian characterize locally stable and unstable directions. After excluding the translational zero modes, the fraction of unstable modes is
\begin{equation}
f_u(T)=\frac{1}{n_{\mathrm{mode}}}\sum_{\alpha=1}^{n_{\mathrm{mode}}}
\Theta\left[-\omega_\alpha^2(T)\right].
\label{eq:fu}
\end{equation}
Here, $n_{\mathrm{mode}}=dN-d$ is the number of nontranslational modes. We define $T_{\mathrm{INM}}$ as the temperature at which the low-temperature linear extrapolation of $f_u^2(T)$ vanishes. As shown in Fig.~\ref{fig:fu_inm}(a), this linear extrapolation vanishes at a characteristic temperature $T_{\mathrm{INM}}$ that lies distinctly above the ideal glass transition scale, $T_{\mathrm{INM}}>T_{\mathrm{K}} \simeq T_\lambda$. The evolution of the corresponding INM spectra is shown in Fig.~\ref{fig:fu_inm}(b): while the spectral weight at $\omega^2<0$ is progressively depleted upon cooling, it remains finite across all accessible equilibrium temperatures. 

The ordering $T_{\mathrm{INM}}>T_\lambda$ is consistent with chaos persisting without negative instantaneous modes, since temporal curvature fluctuations can sustain parametric instability even in a locally stable landscape~\cite{Riemannian_1996_54_5969}. This interpretation concerns extrapolated temperatures; negative modes remain present throughout the accessible equilibrium range.

\FloatBarrier


\begin{thebibliography}{39}

\bibitem{Supercooled_2001_410_259a}
P.~G. Debenedetti and F.~H. Stillinger, Supercooled liquids and the glass transition, \href{https://doi.org/10.1038/35065704}{Nature \textbf{410}, 259 (2001)}.

\bibitem{Facets_2016_69_40}
L.~Berthier and M.~D. Ediger, Facets of glass physics, \href{https://doi.org/10.1063/PT.3.3052}{Physics Today \textbf{69}, 40 (2016)}.

\bibitem{Theoretical_2011_83_587a}
L.~Berthier and G.~Biroli, Theoretical perspective on the glass transition and amorphous materials, \href{https://doi.org/10.1103/RevModPhys.83.587}{Reviews of Modern Physics \textbf{83}, 587 (2011)}.

\bibitem{nature_1948_43_219b}
W.~Kauzmann, The nature of the glassy state and the behavior of liquids at low temperatures, \href{https://doi.org/10.1021/cr60135a002}{Chemical Reviews \textbf{43}, 219 (1948)}.

\bibitem{Configurational_2019_150_160902}
L.~Berthier, M.~Ozawa, and C.~Scalliet, Configurational entropy of glass-forming liquids, \href{https://doi.org/10.1063/1.5091961}{The Journal of Chemical Physics \textbf{150}, 160902 (2019)}.

\bibitem{kauzmann_2023__203}
C.~Cammarota, M.~Ozawa, and G.~Tarjus, The Kauzmann transition to an ideal glass phase, in \textit{Spin Glass Theory and Far beyond: Replica Symmetry Breaking after 40 Years} (World Scientific, 2023) pp.~203--218.

\bibitem{Nature_1958_28_373}
J.~H. Gibbs and E.~A. DiMarzio, Nature of the glass transition and the glassy state, \href{https://doi.org/10.1063/1.1744141}{The Journal of Chemical Physics \textbf{28}, 373 (1958)}.

\bibitem{temperature_1965_43_139a}
G.~Adam and J.~H. Gibbs, On the temperature dependence of cooperative relaxation properties in glass-forming liquids, \href{https://doi.org/10.1063/1.1696442}{The Journal of Chemical Physics \textbf{43}, 139 (1965)}.

\bibitem{Meanfield_2010_82_789}
G.~Parisi and F.~Zamponi, Mean-field theory of hard sphere glasses and jamming, \href{https://doi.org/10.1103/RevModPhys.82.789}{Reviews of Modern Physics \textbf{82}, 789 (2010)}.

\bibitem{Glass_2017_8_265}
P.~Charbonneau, J.~Kurchan, G.~Parisi, P.~Urbani, and F.~Zamponi, Glass and jamming transitions: From exact results to finite-dimensional descriptions, \href{https://doi.org/10.1146/annurev-conmatphys-031016-025334}{Annual Review of Condensed Matter Physics \textbf{8}, 265 (2017)}.

\bibitem{Theory_2007_58_235}
V.~Lubchenko and P.~G. Wolynes, Theory of structural glasses and supercooled liquids, \href{https://doi.org/10.1146/annurev.physchem.58.032806.104653}{Annual Review of Physical Chemistry \textbf{58}, 235 (2007)}.

\bibitem{Colloquium_2015_87_183}
T.~R. Kirkpatrick and D.~Thirumalai, Colloquium: Random first order transition theory concepts in biology and physics, \href{https://doi.org/10.1103/RevModPhys.87.183}{Reviews of Modern Physics \textbf{87}, 183 (2015)}.

\bibitem{Complex_2008_143_a}
W.~G\"otze, \textit{Complex Dynamics of Glass-Forming Liquids: A Mode-Coupling Theory}, Vol.~143 (OUP Oxford, 2008).

\bibitem{Dynamics_2010_61_191a}
D.~Chandler and J.~P. Garrahan, Dynamics on the way to forming glass: Bubbles in space-time, \href{https://doi.org/10.1146/annurev.physchem.040808.090405}{Annual Review of Physical Chemistry \textbf{61}, 191 (2010)}.

\bibitem{Chaos_1998__}
P.~Gaspard, \textit{Chaos, Scattering and Statistical Mechanics} (Cambridge University Press, Cambridge, 1998).

\bibitem{Chaos_2002__}
E.~Ott, \textit{Chaos in Dynamical Systems}, 2nd ed. (Cambridge University Press, Cambridge, 2002).

\bibitem{Probleme_1907_2eserie9_203}
A.~Liapounoff, Probl\`eme g\'en\'eral de la stabilit\'e du mouvement, \href{https://doi.org/10.5802/afst.246}{Annales de la Facult\'e des sciences de Toulouse : Math\'ematiques \textbf{2e s\'erie, 9}, 203 (1907)}.

\bibitem{Viscous_1969_51_3728b}
M.~Goldstein, Viscous liquids and the glass transition: A potential energy barrier picture, \href{https://doi.org/10.1063/1.1672587}{The Journal of Chemical Physics \textbf{51}, 3728 (1969)}.

\bibitem{topographic_1995_267_1935a}
F.~H. Stillinger, A topographic view of supercooled liquids and glass formation, \href{https://doi.org/10.1126/science.267.5206.1935}{Science \textbf{267}, 1935 (1995)}.

\bibitem{Potential_2005_2005_P05015a}
F.~Sciortino, Potential energy landscape description of supercooled liquids and glasses, \href{https://doi.org/10.1088/1742-5468/2005/05/P05015}{Journal of Statistical Mechanics: Theory and Experiment \textbf{2005}, P05015 (2005)}.

\bibitem{Lyapunov_1980_15_9}
G.~Benettin, L.~Galgani, A.~Giorgilli, and J.-M. Strelcyn, Lyapunov characteristic exponents for smooth dynamical systems and for Hamiltonian systems; a method for computing all of them. Part 1: Theory, \href{https://doi.org/10.1007/BF02128236}{Meccanica \textbf{15}, 9 (1980)}.

\bibitem{Lyapunov_1980_15_21}
G.~Benettin, L.~Galgani, A.~Giorgilli, and J.-M. Strelcyn, Lyapunov characteristic exponents for smooth dynamical systems and for Hamiltonian systems; a method for computing all of them. Part 2: Numerical application, \href{https://doi.org/10.1007/BF02128237}{Meccanica \textbf{15}, 21 (1980)}.

\bibitem{Extensivity_2017_95_22102}
M.~Das, A.~B. Costa, and J.~R. Green, Extensivity and additivity of the Kolmogorov-Sinai entropy for simple fluids, \href{https://doi.org/10.1103/PhysRevE.95.022102}{Physical Review E \textbf{95}, 022102 (2017)}.

\bibitem{chaotic_2013_9_288a}
E.~J. Banigan, M.~K. Illich, D.~J. Stace-Naughton, and D.~A. Egolf, The chaotic dynamics of jamming, \href{https://doi.org/10.1038/nphys2593}{Nature Physics \textbf{9}, 288 (2013)}.

\bibitem{Critical_2019_10_2155}
M.~Das and J.~R. Green, Critical fluctuations and slowing down of chaos, \href{https://doi.org/10.1038/s41467-019-10040-3}{Nature Communications \textbf{10}, 2155 (2019)}.

\bibitem{Lyapunov_1996_76_3738}
S.~Sastry, Lyapunov spectra, instantaneous normal mode spectra, and relaxation in the Lennard-Jones liquid, \href{https://doi.org/10.1103/PhysRevLett.76.3738}{Physical Review Letters \textbf{76}, 3738 (1996)}.

\bibitem{Lyapunov_1996_230_364}
Ch.~Dellago and H.~A. Posch, Lyapunov instability, local curvature, and the fluid-solid phase transition in two-dimensional particle systems, \href{https://doi.org/10.1016/0378-4371(96)00069-6}{Physica A: Statistical Mechanics and its Applications \textbf{230}, 364 (1996)}.

\bibitem{Curvature_1997_56_2508}
V.~Mehra and R.~Ramaswamy, Curvature fluctuations and the Lyapunov exponent at melting, \href{https://doi.org/10.1103/PhysRevE.56.2508}{Physical Review E \textbf{56}, 2508 (1997)}.

\bibitem{Crossover_2024_37_95402}
C.~Cockrell and R.~W. Grimes, Crossover in atomic mobility underlying the glass transition in inorganic glasses, \href{https://doi.org/10.1088/1361-648X/ad9fc8}{Journal of Physics: Condensed Matter \textbf{37}, 095402 (2024)}.

\bibitem{Hioe_1987_35_847}
F.~T. Hioe and Z.~Deng, Stability-instability transitions in Hamiltonian systems of \textit{n} dimensions, \href{https://doi.org/10.1103/PhysRevA.35.847}{Physical Review A \textbf{35}, 847 (1987)}.

\bibitem{Kinetic_2000_101_231}
R.~Van~Zon, H.~Van~Beijeren, and J.~R. Dorfman, Kinetic theory estimates for the Kolmogorov--Sinai entropy, and the largest Lyapunov exponents for dilute, hard ball gases and for dilute, random Lorentz gases, in \href{https://doi.org/10.1007/978-3-662-04062-1_10}{\textit{Hard Ball Systems and the Lorentz Gas}}, Vol.~101 (Springer Berlin Heidelberg, Berlin, Heidelberg, 2000) pp.~231--278.

\bibitem{Configurational_2026_165_084502}
J.~Luo, L.~Liu, J.~F. Douglas, and T.~Sinno, Configurational entropy as a potential metric for empirical potential development and validation, \href{https://doi.org/10.1063/5.0346187}{The Journal of Chemical Physics \textbf{165}, 084502 (2026)}.

\bibitem{Geometric_2000_337_237}
L.~Casetti, M.~Pettini, and E.~G.~D. Cohen, Geometric approach to Hamiltonian dynamics and statistical mechanics, \href{https://doi.org/10.1016/S0370-1573(00)00069-7}{Physics Reports \textbf{337}, 237 (2000)}.

\bibitem{Riemannian_1996_54_5969}
L.~Casetti, C.~Clementi, and M.~Pettini, Riemannian theory of Hamiltonian chaos and Lyapunov exponents, \href{https://doi.org/10.1103/PhysRevE.54.5969}{Physical Review E \textbf{54}, 5969 (1996)}.

\bibitem{JAXMD_2020_33_11479a}
S.~S. Schoenholz and E.~D. Cubuk, JAX M.D. A framework for differentiable physics, in \href{https://papers.nips.cc/paper/2020/file/83d3d4b6c9579515e1679aca8cbc8033-Paper.pdf}{\textit{Adv. Neural Inf. Process. Syst.}}, Vol.~33 (Curran Associates, Inc., 2020).

\bibitem{Understanding_2024_128_10999}
Q.-L. Yuan, X.~Xu, J.~F. Douglas, and W.-S. Xu, Understanding relaxation in the Kob--Andersen liquid based on entropy, string, shoving, localization, and parabolic models, \href{https://doi.org/10.1021/acs.jpcb.4c04806}{The Journal of Physical Chemistry B \textbf{128}, 10999 (2024)}.

\bibitem{glassy_2024_161_34504}
S.~Nath and S.~Sengupta, Is the glassy dynamics same in 2D as in 3D? The Adam--Gibbs relation test, \href{https://doi.org/10.1063/5.0174563}{The Journal of Chemical Physics \textbf{161}, 034504 (2024)}.

\bibitem{Scaling_2004_16_L489}
C.~De~Michele, F.~Sciortino, and A.~Coniglio, Scaling in soft spheres: Fragility invariance on the repulsive potential softness, \href{https://doi.org/10.1088/0953-8984/16/45/L01}{Journal of Physics: Condensed Matter \textbf{16}, L489 (2004)}.

\bibitem{Influence_2024_128_9889}
Q.-L. Yuan, X.~Xu, J.~F. Douglas, and W.-S. Xu, Influence of density and pressure on glass formation in the Kob--Andersen model, \href{https://doi.org/10.1021/acs.jpcb.4c04721}{The Journal of Physical Chemistry B \textbf{128}, 9889 (2024)}.

\end{thebibliography}
\end{document}